# Synthesis and physical properties of a new layered ferromagnet, $Cr_{1.21}Te_2$

Zhixue Shu[1], Haozhe Wang[2], Na Hyun Jo[3], Chris Jozwiak[4], Aaron Bostwick[4], Eli Rotenberg[4], Weiwei Xie[2], Tai Kong[1,5]

*1 Department of Physics, University of Arizona, Tucson, AZ 85721*

*2 Department of Chemistry, Michigan State University, East Lansing, MI 48824*

*3 Department of Physics, University of Michigan, Ann Arbor, MI 48109*

*4 Advanced Light Source, Lawrence Berkeley National Laboratory, Berkeley, CA 94620, USA*

*5 Department of Chemistry and Biochemistry, University of Arizona, Tucson, Arizona, 85721*

Abstract

Single crystals of a new layered compound, $Cr_{1.21}Te_2$, was synthesized via a vapor transport method. The crystal structure and physical properties were characterized by single crystal and powder x-ray diffraction, temperature- and field-dependent magnetization, zero-field heat capacity and angle-resolved photoemission spectroscopy. $Cr_{1.21}Te_2$, containing two Cr sites, crystalizes in a trigonal structure with a space group P-3 (No. 147). The Cr site in the interstitial layer is partially occupied. Physical property characterizations indicate that $Cr_{1.21}Te_2$ is metallic with hole pockets at the Fermi energy and undergoes a ferromagnetic phase transition at ~173 K. The magnetic moments align along the *c*-axis in the ferromagnetic state. Based on low temperature magnetization, the spin stiffness constant $D$ and spin excitation gap $\Delta$ were estimated according to Bloch's law to be $D = 99 \pm 24$ meV Å$^2$ and $\Delta = 0.46 \pm 0.33$ meV, suggesting its possible application as a low dimensional ferromagnet.

## I. Introduction

Since the discovery of intrinsic long range ferromagnetism in two-dimensional van der Waals materials [1,2], there has been a growing interest in the discovering, synthesis and physical property investigation of layered ferromagnets, aiming to explore possible low power spintronic device application and fundamental research in low-dimensional magnetism[3,4] .

Among all known layered compounds, Cr based materials attracted wide research interest as many of them are ferromagnets [1,2,4–8]. Cr-Te binary phases, in particular, host a wide range of chemical structure, electronic, and magnetic properties which can offer perpendicular magnetic anisotropy[9] and tunability of properties[10] depending on the Cr concentration. Theoretical calculations predict that binary chromium chalcogenides have the potential to be a good candidate for two-dimensional magnets[11,12]. However, despite decades of research, the Cr-Te binary phase space is not yet fully uncovered. Depending on the synthesis condition and chromium content, the general chemical formular, $Cr_{1+\delta}Te_2$, can take different crystal structure. Previous investigations on various compositions, examples being bulk CrTe[13,14], $Cr_{23}Te_{24}$[14], $Cr_{0.95}Te$[15], $Cr_7Te_8$[14,16], $Cr_5Te_6$[17], $Cr_4Te_5$[18], $Cr_3Te_4$[19], $Cr_2Te_3$[20–21], $Cr_5Te_8$[22–27], $Cr_{9.35}Te_{16}$ [28], $Cr_{4.18}Te_8$[29], $CrTe_2$[30,31] and various thin films[32,33], show that they have ferromagnetic transitions with Curie temperatures ranging from ~160 K – 360 K. $Cr_{1+\delta}Te_2$ compounds share similar crystal structures. Along the *c*-axis of the structure, there is alternating stacking of a fully occupied Cr layer and a Cr vacancy layer. Cr atoms in fully occupied layers form hexagonal structure and are surrounded by Te anions, forming edge-sharing octahedra. Excessive Cr atoms are intercalated between $CrTe_2$ layers with different number of vacancies depending on the Cr content. At high Cr concentration, the crystal structure becomes more NiAs-type like, as in $Cr_5Te_6$[17] and $Cr_7Te_8$[14,16]. Therefore, the composition of all these compounds can be conveniently represented by the notion $Cr_{1+\delta}Te_2$ where δ is the fraction of Cr atoms self-intercalated between neighboring $CrTe_2$ layers. Electronically, with different Cr content and structure, these Cr-Te binary compounds could be semiconducting, half-metallic or metallic[28,34,35]. $CrTe_3$ is the only reported Cr-Te binary compound with antiferromagnetic coupling with a Neel temperature about 55 K[35].

In exploring new ferromagnetic material in the Cr-Te phase space, here we report the single crystal growth, crystal structure, magnetic, electronic and thermodynamic properties of a new Cr-Te binary compound $Cr_{1.21}Te_2$.

## II. Experimental methods

$Cr_{1.21}Te_2$ single crystals were obtained when trying to synthesize $Cr_2Te_3$ by using the chemical vapor transport method. Starting elements (chromium powder, Alfa Aesar, 99.94%; tellurium lump, Alfa Aesar, 99.999%) with an atomic ratio of Cr:Te = 2:3 were sealed in a silica tube under vacuum with ~3.2 mg/ml iodine (Alfa Aesar, 99.8%) serving as vapor transport agent. The silica tube was then placed in an open-ended one-zone tube furnace. The material side was slowly heated up to 850 °C, while the cold end was about 100 degrees lower in temperature. The tube was kept at 850 °C for a day before slowly cooled to 750 °C at a rate of 1 °C/min. The tube was then kept there for 7 days before air quenching to room temperature. Plate-like $Cr_{1.21}Te_2$ single crystals were found at the cold end of the silica tube, which can be exfoliated by the conventional scotch tape method. Representative single crystals are shown in Figure 1 on a millimeter grid paper.

The obtained single crystals were ground into power for powder x-ray diffraction (PXRD) measurement. PXRD was conducted using a Bruker D8 Discover diffractometer with a microfocus and Cu $K_\alpha$ radiation (λ = 1.54 Å). PXRD data were analyzed using GSAS and the Le Bail method[36,37]. Single crystal x-ray diffraction (SCXRD) was conducted using a Bruker D8 Quest Eco single crystal x-ray diffractometer, equipped with Mo radiation (λ = 0.71073 Å) with an $\omega$ of 2.0° per scan and an exposure time of 10 s per frame. A SHELXTL package with the direct methods and full-matrix least-squares on the $F^2$ model was used to determine the crystal structure of Cr-Te.

Magnetization and heat capacity were measured using a quantum design physical property measurement system (PPMS) Dynacool (1.8 K - 300 K, 0 - 90 kOe). Temperature- and field-dependent magnetization was measured using a vibrating-sample magnetometer. Heat capacity was measured using the relaxation method. Angle-resolved photoemission spectroscopy (ARPES) experiments were conducted at Beamline 7.0.2 (Maestro), Advanced Light Source. Samples were cleaved in situ by knocking off an alumina post attached to the single crystals of $Cr_{1.21}Te_2$. The beam diameter was ~ 15 μm, and the measurement temperature was ~ 11 K.

Electronic structure calculation was conducted on a hypothetical "$CrTe_2$" instead of $Cr_{1.21}Te_2$ using Vienna Ab initio Simulation Package (VASP)[38] version 5.4.4, which implements the projector augmented wave (PAW)[39] method. The structure was first taken from the refined crystallographic data and then relaxed until the maximum absolute total force on each atom was smaller than $10^{-3}$ eV/Å and the energy converged to $10^{-8}$ eV. The generalized gradient approximation (GGA) parameterized Perdew-Burke-Ernzerhof (PBE) functional[40] as used to account for the electronic exchange and correlation. A kinetic energy cutoff of 800 eV was used for the plane-wave basis set. Γ-point centered Monkhorst–Pack[41] k-point grids of 18 × 18 × 8 and 33 × 33 × 16 were applied to sample the Brillouin zone for self-consistent field calculation and Fermi surface. VASPKIT[42] and IFermi[43] package were used for data processing.

## III. Results and discussion

$Cr_{1.21}Te_2$ crystalizes in trigonal structure with the space group P-3 (No. 147), determined by SCXRD refinement. The possible superlattice choices were carefully examined and excluded, including monoclinic $Cr_5Te_8$, hexagonal CrTe and $Cr_2Te_3$. The P-3m (#164) was also used to refine the structure. The highest peaks and deepest holes in P-3m (~ +13 and -10 e-/Å$^3$) are much larger than the ones from the refinement using P-3 (~ +5.3 and -7.5 e-/Å$^3$). Moreover, the $R_1$ and w$R_2$ are also significantly larger according to the Hamilton test when using P-3m. Detailed refinement results of $Cr_{1.21}Te_2$ are listed in Tables 1 and 2.

The refined crystal structure of $Cr_{1.21}Te_2$ is shown in Figure 1 (a)-(b) viewing along and perpendicular to the *c*-axis. Cr and Te atoms are represented by blue and golden spheres. The crystal structure is characterized by alternating layers of Cr and Te, with Cr vacancies occurring in every other Cr layer. Cr ions are each surrounded with six Te atoms forming $CrTe_6$ octahedra. The crystal structure of $Cr_{1.21}Te_2$ can be viewed as the stacking of edge-sharing $CrTe_2$ octahedron layers along the *c*-axis and partially occupied Cr atoms sitting in between these fully occupied $CrTe_2$ layers. In each $CrTe_2$ layer, Cr atoms form a triangular lattice (Figure 1(a)). In the partially occupied interstitial layers, each $CrTe_6$ octahedron is connected to neighboring octahedra along the *c*-axis via face-sharing [21]. Figure 1 (c) shows a scanning electron microscopy (SEM) image of a single crystal. Lines are terraces from different layers. Inset shows an optical image of a representative single crystal. The chemical composition was estimated via energy-dispersive x-ray spectroscopy (EDX), and the atomic ratio of Cr and Te is approximately 1.21: 2.0, consistent with the single crystal refinement result.

Figure 1(d) shows the PXRD diffraction result. The observed diffraction data match well with calculated intensities using the LeBail method. The consistency also suggests a good phase purity of synthesized samples. It worth noting that PXRD alone would not be able to distinguish between P-3 and P-3m space groups. Single crystal diffraction is necessary to identify the proper structural model for $Cr_{1.21}Te_2$.

Temperature-dependent magnetic susceptibility ($\chi = M/H$) of $Cr_{1.21}Te_2$ for *H*//*c* is shown in Figure 2(a). At high temperature, the magnetic susceptibility follows the Curie-Weiss behavior. By fitting $\chi$ in the temperature range 220 K – 300 K to the Curie-Weiss law $1/\chi = (T - \theta_{CW})/C$, we obtained the Curie-Weiss temperature $\theta_{CW} =$ 188 K and effective magnetic moment 4.5 $\mu_B$/Cr. The Curie-Weiss temperature and effective moment along *ab* plane are close to the values obtained for *c*-axis, and therefore not shown in the figure. The large positive Curie-Weiss temperature indicates a dominating ferromagnetic exchange interaction in $Cr_{1.21}Te_2$. At low temperature, the magnetic susceptibility rises sharply at around 180 K, corresponding to a paramagnetic to ferromagnetic phase transition. The ferromagnetic transition temperature, $T_c$, was determined by the temperature derivative of the susceptibility, which manifests a sharp peak at 174 K.

Figure 2(b) shows magnetic isotherms measured at 1.8 K for both *H*//*ab* and *H*//*c*. A strong magnetic anisotropy is evidenced by an easy axial magnetic moment and hard in-plane magnetization that saturates at ~80 kOe. The saturation magnetization is about 2.14 $\mu_B$/Cr for both orientations, which is smaller than expected value of 3 $\mu_B$/Cr for $Cr^{3+}$. Although not shown in the figure, a very small hysteresis was observed, indicating the as-grown samples are soft ferromagnets.

The ferromagnetic phase transition is confirmed by heat capacity measurement. Figure 3 shows the temperature-dependent specific heat of $Cr_{1.21}Te_2$ under zero magnetic field. At high temperatures, heat capacity approaches the classical Dulong–Petit limit $C = 3NR = 80.32$ J/mol·K. An λ-peak anomaly, suggesting a second order phase transition, was observed at 173 K, which agrees with the $T_c$ determined from the magnetization measurement. After subtracting the phonon contribution to heat capacity using a polynomial fit around the transition temperature, the magnetic entropy estimated from the λ peak is $\Delta S = \sim 0.45$ J/mol·K. This value is similar to other quasi two-dimensional magnetic systems[30,44], which was argued to come from itinerant magnetism contributions. At low temperatures, $C_p$ can be described by $C_p = \gamma T + \beta T^3$, where $\gamma$ is the Sommerfeld coefficient and $\beta$ term describes the phonon contribution. The inset in Figure 3 shows a linear fit to $\frac{C_p}{T}$ versus $T^2$ plot at low temperatures. The good fit by considering the electron and phonon contribution to heat capacity means that magnon contribution to heat capacity at low temperature is negligibly small. The fitted values are $\gamma \sim 12\text{mJ/mol}\cdot\text{K}^2$ and $\beta = 0.86\ \text{mJ/mol}\cdot\text{K}^4$. The Debye temperature thus obtained is $\theta_D \sim 193.6\ \text{K}$, a value typical of other layered materials [30].

As a new layered ferromagnet, it is of great interest to investigate further the magnetic exchange interaction and magnetic anisotropic energy in $Cr_{1.21}Te_2$. Both energy scales are important for the establishment of long-range magnetic ordering in layered magnets according to the Mermin-Wagner theorem[45]. From bulk magnetization, the spontaneous magnetization at low temperature follows the Bloch's law [46],

$$M(0,H) - M(T,H) = g\mu_B\left(\frac{kT}{4\pi D}\right)^{\frac{3}{2}} f_{\frac{3}{2}}(\Delta'/kT). \qquad (1)$$

where $M\,(0,\ H)$ is the spontaneous magnetization at zero temperature and applied magnetic field $H$; $M\,(T,\ H)$ is the spontaneous magnetization at a temperature $T$ and applied magnetic field $H$; $f_p(y) = \sum_{n=1}^{\infty}\frac{e^{-ny}}{n^p}$ is the Bose-Einstein integral function that contains the field-dependent spin excitation gap $\Delta' = \Delta + g\mu_B H$. $D$ is proportional to exchange interaction[11] and $\Delta$ is the zero-field excitation gap which is closely related to magnetic anisotropy[47].

To find the spontaneous magnetization at each temperature, we performed field-dependent magnetization measurement at low temperatures. Spontaneous magnetization at each temperature was obtained by finding the zero field intercept from a linear fitting at high field range, as described in the previous work[48]. Figure 4 shows extracted spontaneous magnetization as a function of temperature, with different symbols representing measurements from different samples and the color-matching lines are fittings based on Equation 1. The fit parameters are $D = 99 \pm 24$ meV Å$^2$ and $\Delta = 0.46 \pm 0.33$ meV. The uncertainties are calculated from the standard deviations of fitting parameters for the 4 measured samples. Compared to other Cr based van der Waals ferromagnets[48] $CrI_3$, $CrGeTe_3$ and $CrSiTe_3$, $Cr_{1.21}Te_2$ has a much higher spin stiffness, indicating a stronger ferromagnetic exchange interaction which is consistent with its higher Curie temperature. Spin wave excitation gap of $Cr_{1.21}Te_2$ is lower than that in $CrI_3$, but comparable to that in $CrSiTe_3$ and $CrGeTe_3$[48]. As magnetic anisotropy stabilizes long-range magnetic order at low dimensions, $Cr_{1.21}Te_2$ may be comparable to $CrGeTe_3$ as an alternative system for a two-dimensional magnet.

Figure 5 shows band structure calculation on the hypothetical $CrTe_2$ and ARPES results on a single crystal of $Cr_{1.2}Te_2$. Band structures of high symmetry points are shown in Figure 5 (a)-(b) for spin-up and spin-down, respectively. Most of the bands around the Fermi energy belong to the Cr 3$d$ and Te 5$s$, $p$ bands. The spin-polarized band structures reveal that the Cr 3$d$ bands are relatively flat near the Brillouin zone boundaries (near point M and along the direction M-Γ and M-K). These features usually arise from magnetic and layered structural influences on the orbital interactions between Cr and Te atoms. Figure 5 (c)-(d) are calculated Fermi surface cuts at fixed $k_z = 0$, and 0.3$c$. Different colors indicate different bands crossing the Fermi energy, as shown in (a)-(b). Corresponding Fermi surface cuts from the ARPES experiment at the photon energy of 120 eV and 131 eV are presented in Figure 5 (e) and (g). Fermi surfaces from the calculations and ARPES experiments are overall in good agreement. Band structures for each photon energy at $k_y = 0$, projections along the Γ-M direction, are plotted in Figure 5 (f) and (h). Considering the band structure calculation results shown in Figure 5 (a) and (b), the chemical potential is shifted ~1 eV upward, which is consistent with additional Cr atoms in measured material than the stoichiometry used in the calculation.

Going back to the magnetic property, it worth noting that the effective magnetic moment of $Cr_{1.21}Te_2$ is relatively large compared to the measured saturation moment. This observation could be partially attributed to the itinerant nature of the magnetism in $Cr_{1.21}Te_2$. The degree of itinerancy can be characterized by the Rhodes-Wohlfarth ratio (RWR), which is defined as $P_c/P_s$. $P_c$ can be obtained from the effective magnetization $P_c(P_c + 2) = P_{\mathrm{eff}}^2$, where $P_{\mathrm{eff}}$ is the effective moment. $P_s$ represents the saturation magnetization[49,50]. For localized magnetic systems RWR is equal to 1 and is larger than unity in itinerant systems. Using values obtained above, the estimated RWR is 1.67 for $Cr_{1.21}Te_2$, similar to that of $Cr_{0.62}Te$[24], thereby indicating weak itinerant ferromagnetism of the material $Cr_{1.21}Te_2$. RWRs for other $Cr_{1+\delta}Te_2$ compounds are shown in Figure 6. Given available data in literatures, there seems to be a broad maximum of RWR ratio at δ ~0.5. The itinerancy of $Cr_{1.21}Te_2$ falls under the overall trend among this series of compounds.

As the effective moment is generally larger than the measured saturation moment in $Cr_{1+\delta}Te_2$, other causes have been investigated besides the itinerant contribution [21,23]. Possible reasons include different magnetic moments on different Cr sites or canted antiferromagnetic components [15,22–24,26,33,51–53]. Cr ions in the partially occupied layers can have nearly negligible magnetic moments, which interact antiferromagnetically and align in a direction other than ferromagnetic easy directions[51]. Neutron diffraction measurements of single crystals are needed in the future to determine more precisely the magnetic structure of $Cr_{1.21}Te_2$ for a complete understanding of the lowered saturation magnetization.

Compared to other $Cr_{1+\delta}Te_2$ compounds, the $T_c$ of $Cr_{1.21}Te_2$ is also consistent with the overall trend where larger $\delta$ leads to higher $T_c$ as shown in Figure 6. The physical reason behind this trend is yet to be determined. As more interstitial Cr atoms can be found between layers at higher $\delta$ values, it is possible that interlayer coupling plays a key role in enhancing the $T_c$.

## IV. Conclusion

In summary, we report the synthesis and physical properties characterization of layered $Cr_{1.21}Te_2$. Being a new compound in the Cr-Te binary, $Cr_{1.21}Te_2$ crystallizes with a space group P-3, determined by single crystal refinement. Its crystal structure can be understood as fully occupied $CrTe_2$ layers with partially occupied Cr sites in between layers. Magnetization measurements suggest a ferromagnetic phase transition at ~173 K with an easy axial moment in the ordered state. Through fitting low temperature spontaneous magnetization using Bloch's law, the spin stiffness and spin excitation gap are estimated to be $D = 99 \pm 24$ meV Å$^2$ and $\Delta = 0.46 \pm 0.33$ meV. The sizable spin excitation gap, as compared to other Cr-based layered ferromagnet, suggests $Cr_{1.21}Te_2$ may sustain its long-range ferromagnetism at two-dimensional limit. Band structure calculation and measurement suggest a metallic state with a hole pocket at the $\Gamma$ point.

## V. Acknowledgements

H.W. and W.X. was supported by U.S. DOE-BES under Contract DE SC0022156. H.W. and W.X. thank the computing resource support from Rutgers University Parallel Computing (RUPC), Center of Materials Theory, Department of Physics and Astronomy. N.H.J was supported by QSA, funded by the U.S. Department of Energy, Office of Science, National Quantum Information Science Research Centers. This research used resources of the Advanced Light Source, which is a DOE Office of Science User Facility under contract no. DE-AC02-05CH11231. Work at the University of Arizona was supported by University of Arizona startup fund.

**Table 1.** Single crystal crystallographic data and structure refinement for $Cr_{1.21(3)}Te_2$.

| Refined Formula | $Cr_{1.21(3)}Te_2$ |
|---|---|
| Space group, $Z$ | $P$-3 (#147), 1 |
| $a$ (Å) | 3.915(2) |
| $c$ (Å) | 6.022(2) |
| V (Å$^3$) | 79.93(9) |
| Absorption correction | Multi-scan |
| Extinction coefficient | None |
| θ range (°) | 3.383 – 34.94 |
| $hkl$ ranges | $-6 \leq h, l \leq 6$<br>$-9 \leq k \leq 9$ |
| No. reflections, $R_{int}$ | 562, 0.0417 |
| No. independent reflections | 238 |
| No. parameters | 9 |
| $R_1$, $wR_2$ (all $I$) | 0.0735, 0.1515 |
| Goodness of fit | 1.394 |
| Highest peaks and deepest holes (e-/Å$^3$) | 5.351; -7.550 |

**Table 2.** Atomic coordinates and isotropic displacement parameters of $Cr_{1.21(3)}Te_2$. ($U_{eq}$ is defined as one-third of the trace of the orthogonalized $U_{ij}$ tensor.)

| Atom | Wyck. | Occ. | $x$ | $y$ | $z$ | $U_{eq}$ |
|---|---|---|---|---|---|---|
| Cr1 | 1$b$ | 1 | 0 | 0 | ½ | 0.011(1) |
| Cr2 | 1$a$ | 0.21(3) | 0 | 0 | 0 | 0.004(7) |
| Te | 2$d$ | 1 | 1/3 | 2/3 | 0.2483(3) | 0.0068(4) |

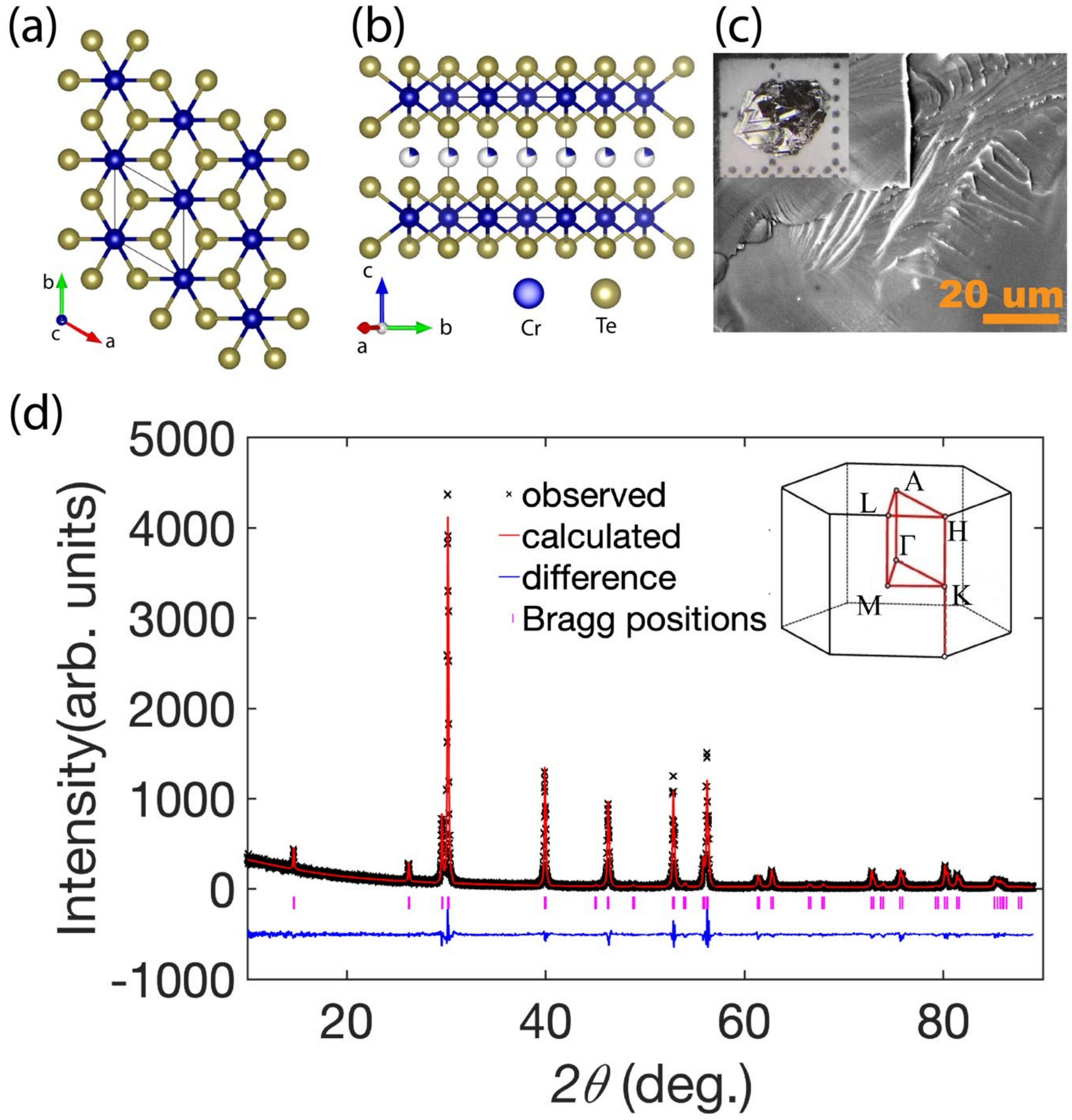

**Figure 1** (a, b) Crystal structure of $Cr_{1.21}Te_2$ (top and side views). Grey lines represent a unit cell. (c) SEM image of a $Cr_{1.21}Te_2$single crystal. Inset: a single crystal of $Cr_{1.21}Te_2$ on a millimeter grid paper. (d) powder x-ray diffraction data of $Cr_{1.21}Te_2$. Inset shows the Brillouin zone.

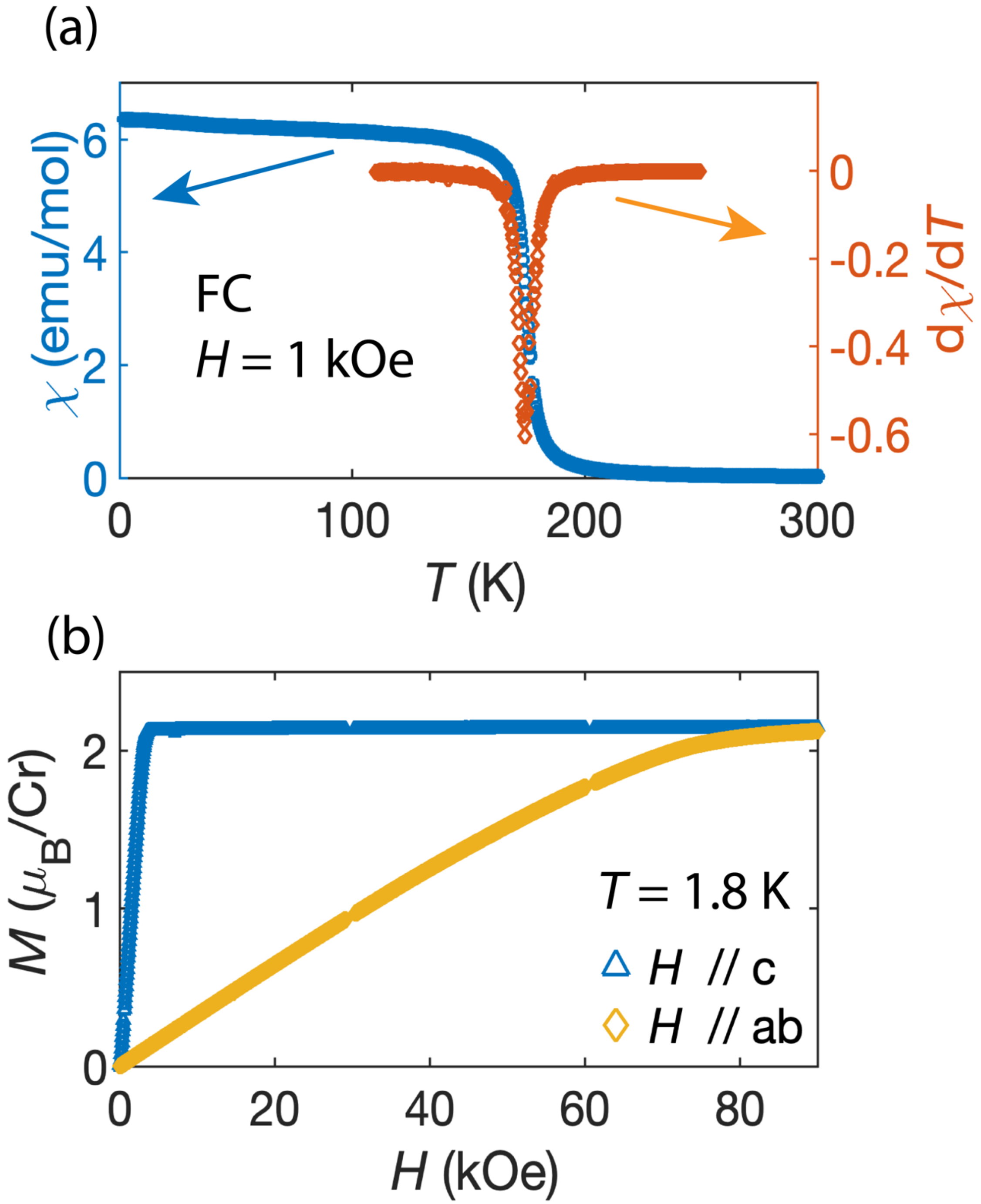

**Figure 2** (a) Temperature-dependent magnetic susceptibility measured at $H$ = 1 kOe on the left y axis and its temperature derivative on the right, (b) Field-dependent magnetization for $H$//c (yellow triangles) and $H$//ab (blue diamonds) measured at 1.8 K.

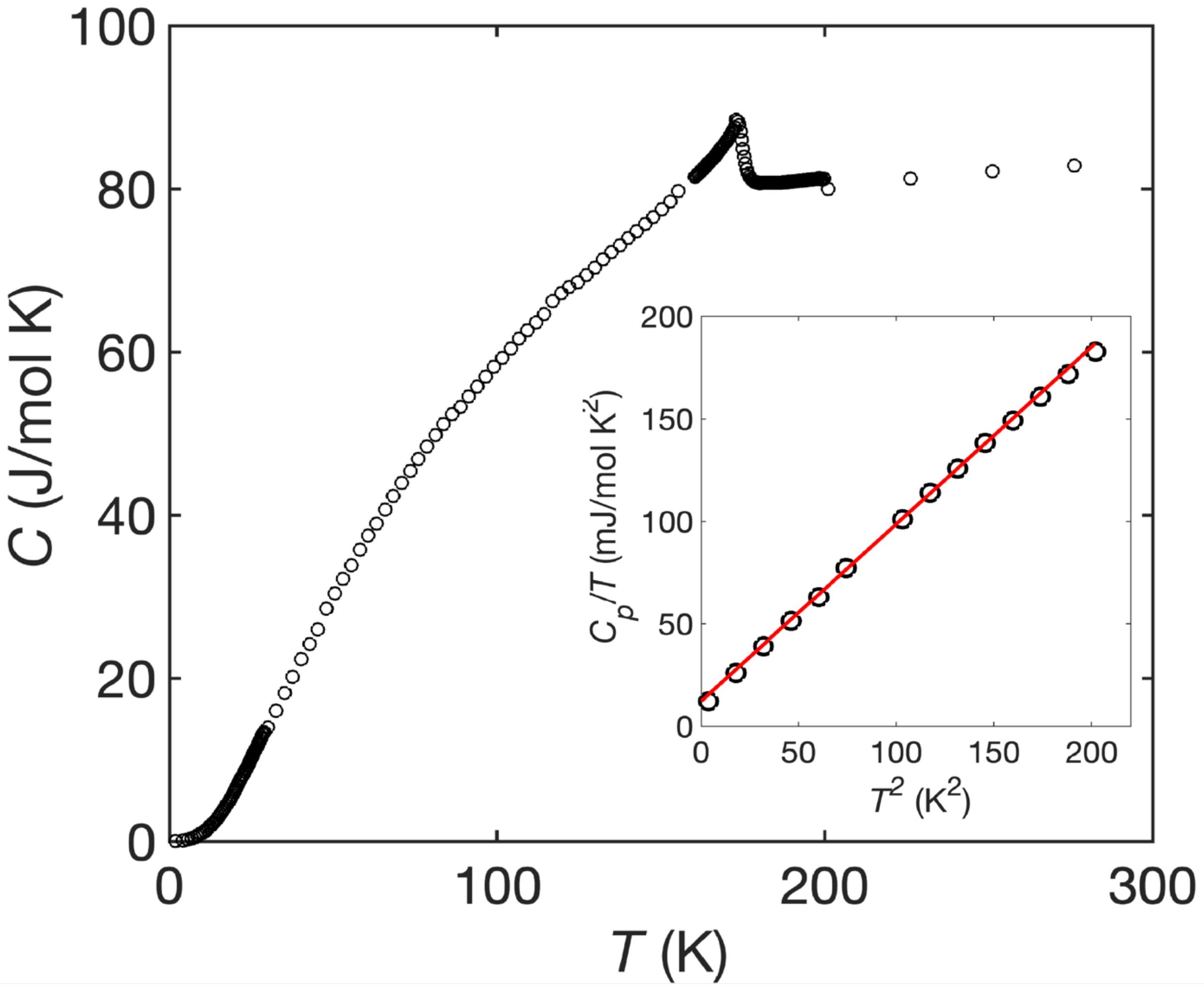

**Figure 3** Zero-field, temperature-dependent specific heat for $Cr_{1.21}Te_2$ Inset shows $\frac{C_p}{T}$ versus $T^2$ at low temperatures. Blue line represents a linear fit.

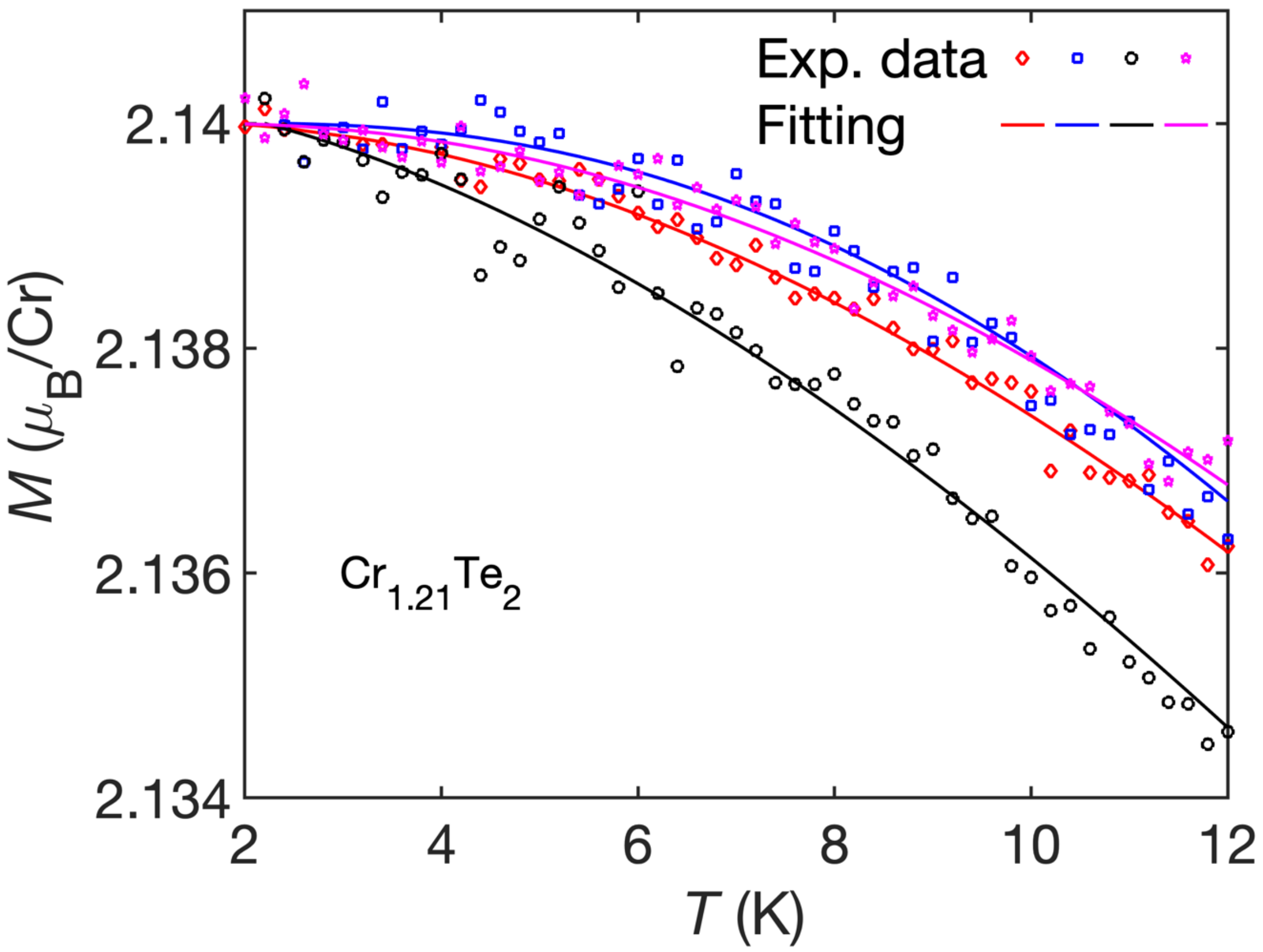

**Figure 4** Temperature-dependent spontaneous magnetization of $Cr_{1.21}Te_2$. Data from different samples are indicated by different colors and symbols. Discreet symbols are experimental data points and color-matching lines are corresponding fittings.

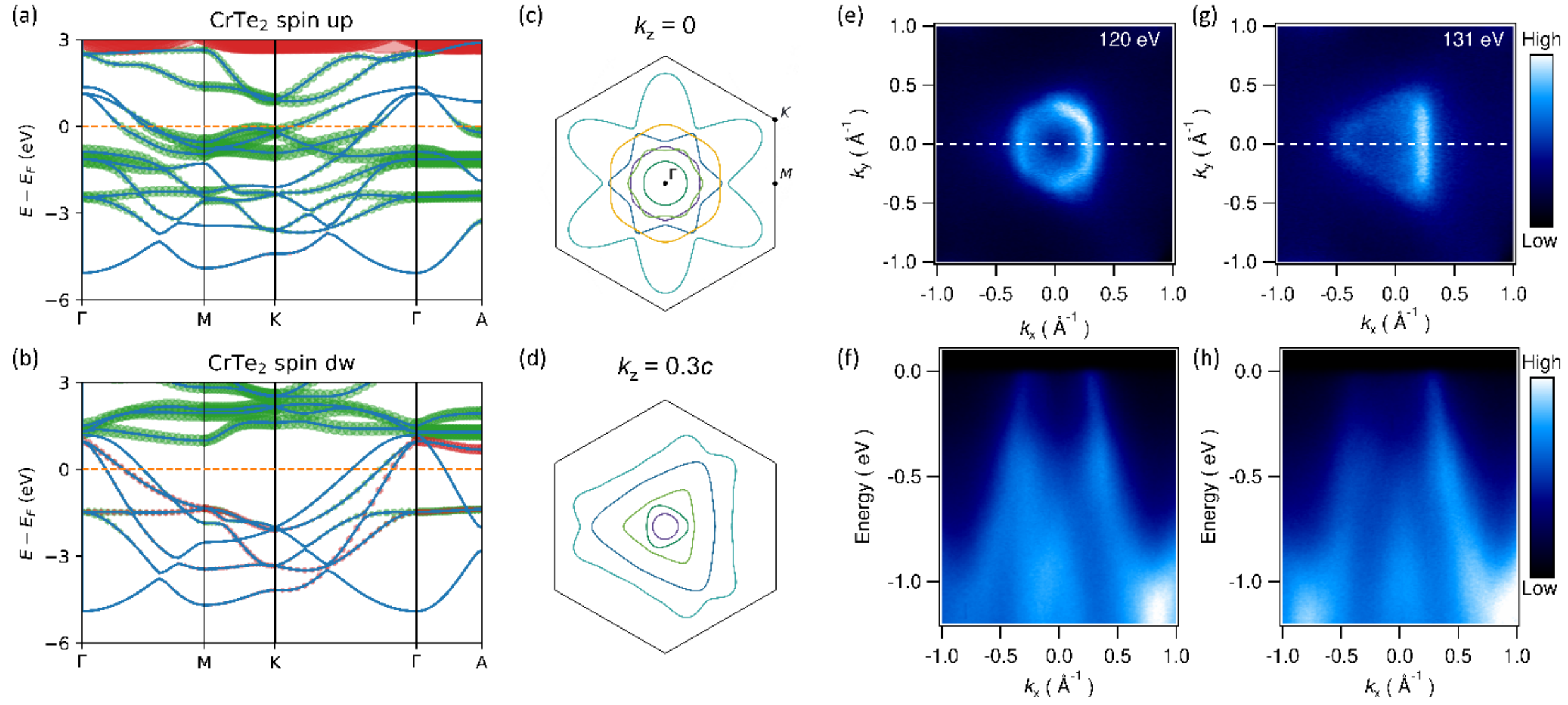

**Figure 5** (a)-(b) Calculated spin polarized band structure of $CrTe_2$ spin up and spin down respectively. Fermi level is highlighted by the orange line. Cr 3*d* and Te 5*s*, 5*p* contribution are shown in green and red, respectively. (c)-(d) Two-dimension slice of the Fermi surface with both spin up and spin down along (001) passing through the Γ point and with a distance 0.3*c* to the Γ point, the $k_x$ from Γ to M is 0.9 $Å^{-1}$ (e) Experimental Fermi surface plots of $Cr_{1.21}Te_2$ at photon energy of 120 eV. (f) Band dispersions along the high symmetry lines of $k_y = 0$ of (e). (g) Fermi surface plots of $Cr_{1.21}Te_2$ at photon energy of 131 eV. (h) Band dispersions along the high symmetry lines of $k_y = 0$ of (g).

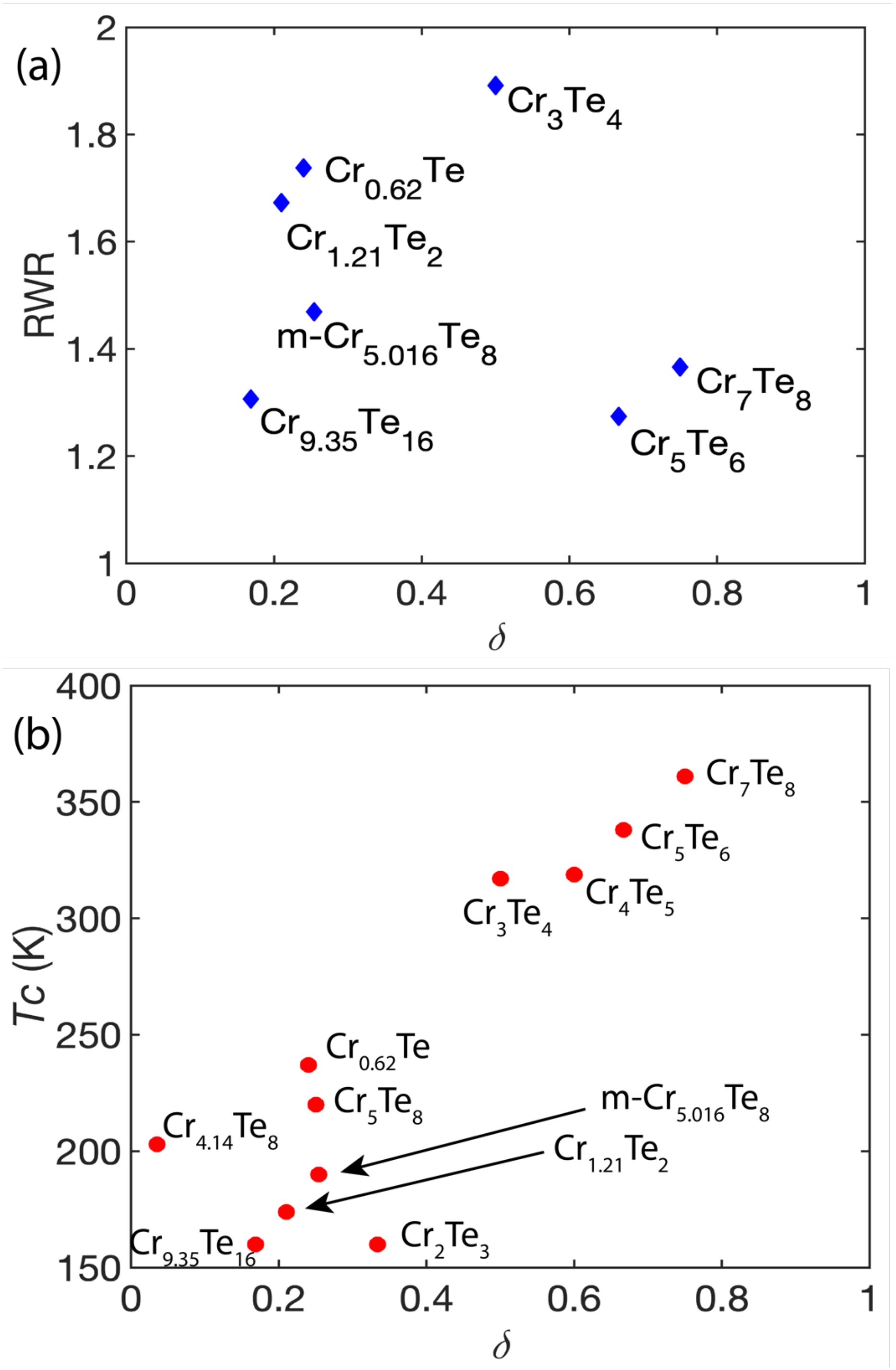

**Figure 6** (a) and (b) Curie temperature and Rhodes-Wohlfarth ratio (RWR) as a function of $\delta$ for binary compounds $Cr_{1+\delta}Te_2$. Data from the following compounds are shown: $Cr_{0.95}Te$[15], $Cr_7Te_8$[14,16], $Cr_5Te_6$[17], $Cr_4Te_5$[18], $Cr_3Te_4$[19], $Cr_2Te_3$[20–21], $Cr_5Te_8$[26], $Cr_{9.35}Te_{16}$[28], $Cr_{4.18}Te_8$[29], m-$Cr_{5.016}Te_8$ [23] and $Cr_{0.62}Te$[54].